\documentstyle[12pt,aasms4]{article}
\def\etal   {{et~al.}\ }

\def\msun{{\rm\,M_\odot}}

\def\vol#1  {{{#1}{\rm,}\ }}

\def\etal{et al.\ }

\def\Sec{\S }

\def\cf{{cf.}\ }

\def\clock{\count0=\time \divide\count0 by 60
     \count1=\count0 \multiply\count1 by -60 \advance\count1 by \time
     \number\count0:\ifnum\count1<10{0\number\count1}\else\number\count1\fi}

\begin{document}
\title{Physical Bias of Galaxies From Large-Scale Hydrodynamic Simulations}
\author{Renyue Cen\altaffilmark{1} and Jeremiah P. Ostriker\altaffilmark{2}}
\altaffiltext{1} {Princeton University Observatory, Princeton University, Princeton, NJ 08544; cen@astro.princeton.edu}
\altaffiltext{2} {Princeton University Observatory, Princeton University, Princeton, NJ 08544; jpo@astro.princeton.edu}

\begin{abstract}
We analyze a new large-scale ($100h^{-1}$Mpc)
numerical  hydrodynamic simulation of the 
popular $\Lambda$CDM cosmological model,
including in our treatment dark matter, gas and
star-formation,
on the basis of standard physical processes.
The method, applied with a numerical resolution
of $<200h^{-1}$kpc (which is still quite coarse for following individual
galaxies, especially in dense regions)
, 
attempts to estimate where and when galaxies form.
We then compare the smoothed
galaxy distribution with the smoothed mass distribution
to determine the ``bias" defined
as $b\equiv (\delta M/M)_{gal}/(\delta M/M)_{total}$
on scales large compared with the code numerical resolution
(on the basis of resolution tests given in the appendix of this
paper).
We find that (holding all variables constant except the quoted one)
bias increases with decreasing scale,
with increasing galactic age or metallicity
and with increasing redshift of observations.
At the $8h^{-1}$Mpc fiducial comoving scale
bias (for bright regions)
is $1.35$ at $z=0$ reaching
to $3.6$ at $z=3$,
both numbers being consistent with extant observations.
We also find that $(10-20)h^{-1}$Mpc voids
in the distribution of luminous objects
are as observed (i.e., observed
voids are not an argument against CDM-like models)
and finally that the younger systems should
show a colder Hubble flow than 
do the early type galaxies (a testable proposition).
Surprisingly, little evolution is found in the amplitude 
of the smoothed 
galaxy-galaxy correlation function (as a function of {\it comoving}
separation).
Testing this prediction vs observations will allow a comparison between
this work and that of Kauffmann \etal which is based
on a different physical modeling 
method.
\end{abstract}


\section{Introduction}

The spatial distribution of galaxies provides the 
core data on which cosmological theories for the growth 
of structure are based.
Spatial correlations,
peculiar velocities, voids, etc.
have provided the primary evidence for the growth of structure due to 
gravitational instabilities.
But the theories developed to model this growth
of structure largely treat the collisionless dark matter,
thought to provide the bulk of the mass density in the universe,
rather than the observables, the stellar parts of galaxies.
Current estimates would put the ratio of the mass densities
of these two components 
at about
($\Omega_{dm}/\Omega_*\approx 0.3/0.003\approx$)
100, so the potential fluctuations 
(except on the smallest scales $\Delta r<10^4$pc)
are dominated by dark matter mass fluctuations.
To accommodate this dichotomy between observables and computables
the concept of ``bias", $b$,
was developed (e.g., Davis \etal 1985) to bridge the gap.
Treated at first as -- what it was --
a convenient way of parameterizing our ignorance,
it has developed
a life of its own.

A common definition would be based on the relation between
the number density of galaxies, $N_g$,
(with, e.g., $|m_V|>|m_{V,0}|$)
or the total mass density in galaxies $\rho_g$
and the total mass in a given region,
$\rho_{tot}$,
with bias,
$b_l$, 
in regions smoothed by a top-hat
smoothing length $l$ defined by
\begin{equation}
<\left({\delta N_g\over N_g}\right)^2>_l\approx
<\left({\delta \rho_g\over \rho_g}\right)^2>_l\equiv b_l^2 <\left({\delta \rho_{tot}\over \rho_{tot}}\right)^2>_l.
\end{equation}
\noindent 
Here we have explicitly put in the scale
dependence,
although in
primitive treatments bias
is sometimes treated as a number.
The realization that $b_l$ is both stochastic and scale dependent
has grown recently
(for discussions of analytic theories of biasing
see Scherrer \& Weinberg 1997 and Pen 1997), but, of course,
this must be the case.
Blanton \etal (1999a) have studied our simulations taking a broader
perspective and allowing, on the right hand side of equation (1),
a dependence on variables other than mass density.
We restrict ourselves in the current exercise to the
conventional parameterization but examine how $b$,
defined by second equality in equation (1),
may depend on spatial scale, galaxy age and other
galactic properties.
Our simulations do not have sufficient spatial
resolution to identify individual galaxies (except
in low density regions where they are
adequate), due to a tendency of the present simulation
to merge distinct
systems in high density regions.
But galaxy mass (if not number)
is conserved during mergers
so the smoothed galaxy density (smoothing scale
$500h^{-1}$kpc)
is the variable we compare with the similarly smoothed dark matter
density.
As this smoothing scale is small compared
to the correlation lengths we examine
($>1h^{-1}$Mpc)
we believe that our numerical resolution is adequate
(but far from ideal)
for the questions examined.
A test, presented in Appendix A,
comparing two simulations with length scale resolution differing
by a factor of two 
reassuringly show no difference in bias,
on the scales studied,
which is greater than statistical error ($\sim 10\%$).

Nature has determined bias,
not by providing us with a mathematically convenient functional
form but through the physical
process of galaxy formation.
By most current accounts the bulk of the baryons
have not condensed into stellar systems,
at the present time
with 
the ratio of gas to stars
$\Omega_b/\Omega_*$ ($\approx 0.04/0.003$) $\ge 10$.
Thus, galaxies appear to have
formed only under 
favorable conditions.
Both simple physical theory and observations
indicate greater formation
efficiencies in regions of higher density,
where cooling processes were more efficient.
In the Galaxy,
the observed ``Schmidt Law"
(Schmidt 1959, also see Kennicutt 1989 for an update) 
stating that star-formation rates scale roughly as the locally
averaged value of $\rho_{gas}^2$ still seems to have
some approximate empirical
validity.
It is based of course on the physical fact that
for the dominant collisionally excited line 
or continuum cooling processes,
the gas cooling rate per unit volume -- thought to be
the rate limiting step --
is proportional to $\rho_{gas}^2$.

Cosmological observations would also seem to support such a picture
since the void regions
are believed to be underdense (from simulations)
by a factor of $3-5$
in total mass density
as compared to the average, but, empirically,
these regions have
a galaxy density 
far below this (Peebles 1993)
in terms of average galaxy density.
Conversely,
in the great clusters of galaxies,
perhaps $10\%$ of the total baryonic mass is in stellar
form, indicating above average efficiency
of galaxy formation in these very overdense regions.

One reasonable 
approach to determining bias is to begin
by noting directly that most stellar galaxies live
within massive halos.
Then, one can compute the distribution of dark matter halos
and estimate which of these will contain which type of stellar
systems.
Much work has been done following this promising
track in recent years using either large
N-body simulations or combining
semi-analytic dark matter treatments
with detailed hydrodynamic simulations 
(Cole \etal 1994; Kauffmann, Nusser \& Steinmetz 1997).
These analyses have produced
results which are consistent with many observations.

However, direct numerical simulations can be made 
on an {\it ab initio} basis which combine
the physical gasdynamic processes used in the hydrodynamic simulations 
of Katz, Hernquist, \& Weinberg (1992)
or Steinmetz (1996) with large scale
numerical simulations
of dark matter pioneered by
Davis, Efstathiou, Frenk and White 
(Davis \etal 1985;
Frenk \etal 1985;
White \etal 1987).
Such calculations could,
in principle,
determine ``bias"
by direct computations.
We have attempted to do that in earlier work (Cen \& Ostriker 1992, 1993b)
but, as is well known, 
the difficulties and uncertainties are formidable.
These are of two kinds:
inadequate physical modeling and insufficient numerical resolution.


In the last five years we have made quite significant improvements in both
our physical modeling and the numerical resolution that we can achieve.
First of all, we have upgraded from
the somewhat diffusive aerospace
gasdynamics-based Eulerian hydrocode (Cen 1992)
to a shock capturing, high order, total variation diminishing (TVD)
hydrocode (Ryu \etal 1993).
A very important additional ingredient in the new TVD code 
is an implementation of a new, entropy variable into the
conventional TVD scheme (Harten 1983).
The true spatial resolution for a given mesh
has increased by a factor about two (Kang \etal 1994)
and,
unphysical spurious heating is removed, 
allowing for a much more accurate treatment in
lower temperature regions.
We have also gained a factor of $2.5$ through Moore's Law
(increasing computational power),
so that the overall gain is approximately a factor of five
from $>1h^{-1}$Mpc to $200h^{-1}$kpc.
In particular, our current simulations have a nominal spatial
resolution which is small compared to 
the typical distance between galaxies (cf. Figure 1),
which was not the case earlier, so,
while we still are totally unable to say anything about the
internal structural properties of galaxies,
we should be able to specify galaxy integral properties: 
we feel that the resolution
suffices to determine where and when 
galaxies of normal range
will form,
and to determine, locally,
the mass density in galaxies while
we remain unable to say very much
about the number density of galaxies.

Secondly, we have substantially improved
the realism of the input physics, aside from a seemingly
better cosmological model.
First, we have added more, relevant microphysical
processes due to elements other than hydrogen and helium.
The inclusion affects directly cooling/heating, optical opacity, radiation
field and indirectly through complex interplays 
among these physical processes and gravity.
Second, in our galaxy formation algorithm, we now allow
not only energy feedback from stars but also
matter (including metals) ejection into the IGM.
Third, we separately but simultaneously 
follow the dynamics of metals in the IGM,
produced during the feedback processes.
The algorithm for identifying regions 
of star-formation which has remained the same has also
been used by other workers making detailed hydrodynamic
simulations:
Katz, Hernquist, \& Weinberg (1992),
Katz, Weinberg, \& Hernquist (1996),
Steinmetz (1996),
Gnedin \& Ostriker (1997) and
Abel \& Norman (1998).
In a region of sufficient overdensity, if three
conditions are {\it simultaneously} satisfied,
we assume that an already existing collapse cannot be
reversed and that collapsed objects of some kind (stars, star
clusters or molecular clouds) will form.
The criteria in detail are as follows:
\begin{eqnarray}
a)& \nabla\cdot{\vec v}<0\nonumber\\
b)& M_{gas} > M_{Jeans}\nonumber\\
c)& t_{cooling} < t_{dynamical}
\end{eqnarray}
\noindent
If all are estimated to be satisfied within a cell,
then a mass in gas is removed from the cell (having gas mass $M_g$)
equal to $\Delta M_g = - M_g \Delta t/t_{dyn}$
in a timestep $\Delta t$ and put into a
collisionless particle of mass 
$\Delta M_* = - M_g$.
That particle is labeled with the time of formation,
the metallicity of the gas from which it was made
and the density of the cell in which is was made.
These ``stellar" particles have typical mass of 
$10^7\msun$ and
can be thought of as clusters within which star-formation
of essentially coeval stars
will occur.
They are followed after birth by the same code that
tracks the collisionless dark matter particles.
We find that (except in high density regions)
these particles group themselves into galaxy like
objects (cf. Fig 1) 
which have a total mass, a mean time of formation (with dispersion)
and a mean metallicity (with dispersion).
Within each ``galaxy" of defined mean age there will be stellar particles
having a wide range of ages as the computed
galaxies (like our own) are assembled over time.
Even in high density regions we can smooth over some spatial
scale and define spatial fields
having ``stellar" density, age and metallicity.

The top (large) panel of Figure 1
shows the dark matter density 
for a random slice of size $100\times 100h^{-2}$Mpc$^2$
and thickness of three cells ($586h^{-1}$kpc).
Each small tickmark of the top panel has a size of $6.25h^{-1}$Mpc.
The three bottom pairs of panels show three galaxy groups in this slice
in an enlarged display; two separate panels are shown for each selected
group with the top panel showing the galaxy density contour
and the bottom panel showing the dark matter density contour.
Each small panel has 
a size of $6.25\times 6.25h^{-2}$Mpc$^2$
and each tickmark for the bottom panels 
has a size of $391h^{-1}$kpc, corresponding to two cells,
which is approximately the resolution of the code
(Cen \& Ostriker 1999).
For Group A  we see that the galaxy particles
have grouped themselves into two quite separated galactic lumps,
whereas the dark matter contour shows only one major lump.
For Group B several galactic lumps with varying mass are seen,
which roughly follow the distribution of the dark matter,
but the difference between galaxy density and dark matter
is noticeable.
In Group C the two (out of four) galactic lumps in the middle
are clearly separated, while they are embedded in a common halo.
Clearly any galaxy pairs that are separated by $1h^{-1}$Mpc
are resolved by the code.
We note the fact that we often see distinct galactic mass objects
swimming separately in a common halo.
This is not due to differences in 
numerical resolution between dark matter 
and galaxy particles:
the galaxy particles and dark matter particles
are followed with the identical PM code.
Rather, this occurs because the galaxy particles formed
from gaseous accumulations that had undergone
radiative cooling and contraction.
In the objects shown as distinct lumps 
all normal dynamical processes are active including
dynamical friction and mergers (which of course
are somewhat overestimated).

Considering the substantial improvements in the code
it seems worthwhile returning to the problem of bias
using new simulations.
The combination of the improvements already noted
allows an identification of individual galaxies
in low density regions.
In addition, some new questions can now be addressed.
For example, we can now study many problems concerning
the distribution of metals, in a self-consistent fashion.
In this paper we study the relative distributions
of galaxy mass density and total mass density, and their evolution in
a cold dark matter model with a cosmological constant ($\Lambda$CDM).
Some aspects (scale dependence of bias)
of this work have been reported on in Blanton \etal (1999a).
Here we focus on other observables including dependence on epoch.

The new simulations
allow us to address the redshift
dependence 
of this bias,
which has also been recently addressed by Katz \etal (1998).
We will also examine the dependence on galaxy mass and metallicity
and the bias of other constituents
such as
uncondensed gas in various
temperature ranges.
Due to the very considerable cost 
of these large scale simulations we have only
completed the analysis of one currently favored
cosmological model.
Future work will allow us to compare bias amongst models.

After describing briefly further properties of the computational method
and the specific cosmological model considered in \Sec 2, 
we present the results in \Sec 3.
\S 3.1 is devoted to galaxy bias at redshift zero,
while \S 3.2 examines the evolution of bias of galaxies with redshift.
Finally, our conclusions are summarized in \Sec 4.
Appendix A presents an examination of the effects of the finite
numerical resolution on our results
by comparing output of two simulations having spatial 
resolution differing by a factor of two.

\section{Method and Cosmological Model}

We model galaxy formation as described earlier.
Self-consistently, 
feedbacks into the IGM from young stars in these galaxy particles
are allowed, in three related forms:
supernova mechanical energy output,
UV photons output
and mass ejection from the same supernova explosions. 

We adopt a cold dark matter model
with a cosmological constant ($\Lambda$CDM),
close to the Concordance Model of Ostriker \& Steinhardt (1995)
and the preferred model of Krauss \& Turner (1995)
with the following parameters: $\Omega=0.37$,
$\Omega_b=0.049$, 
$\Omega_\Lambda=0.63$, $h=0.70$ and $\sigma_8=0.80$.
The model also is consistent with that favored by
recent type Ia supernova observations (Riess \etal 1998).
This model is normalized to both COBE and the observed cluster
abundance at zero redshift (Eke \etal 1996).
The current age of the universe of this model
is $12.7~$Gyrs, consistent with recent 
age constraint from latest globular cluster observations/interpretations
(c.f., Salaris, Degl'Innocenti, \& Weiss 1997).
The choice of the Hubble constant is in agreement
with current observations;
it appears that $H_0(obs)=65\pm 10$km/s/Mpc can account for
the distribution of the current data from various measurements
(see, e.g., Trimble 1997).
The choice of $\Omega_b=0.024h^{-2}$ is consistent
with current observations of Tytler, Fan \& Burles (1996).

The box size of the primary simulation is $100h^{-1}$Mpc and
there are $512^3$ cells (on a uniform mesh),
giving the nominal resolution of $197h^{-1}$kpc.
But actual resolution,
as determined by extensive tests (Cen \& Ostriker 1998),
is approximately of a factor of $1.1-1.7$ worse than this,
depending on the region in question.
There are
$256^3$ dark matter particles used, with 
the mass of each particles being $5.3\times 10^9h^{-1}\msun$.
The typical galaxy particle has a mass of $10^7\msun$.
Besides this primary box,
which will be used for most of our
analysis, we have made a higher
resolution run with box size $50h^{-1}$Mpc,
with the same number of cells,
which will be used for resolution calibration purposes in 
Appendix A. 

\section{Results}

\subsection{Bias of Galaxies At Redshift Zero}

Panel (a) of Figure 2 shows the bias of galaxies
for four sets ($4\times 25\%$ in mass) 
of galaxies ordered by
formation time [z=(0.0-0.8), (0.8-1.0), (1.0-1.3), (1.3-10.0)],
as well as all galaxies at $z=0$,
as a function of top-hat smoothing radius.
All scales depicted are considerably larger
than our resolution limit;
Appendix A indicates that numerical 
resolution problems should not be important
for the results discussed here.
As an important aside we note that $50\%$ of the galaxy mass was
formed between redshifts 0.8 and 1.3, with median galaxy
formation epoch $z=1.0$, consistent with Madau's (1997)
analysis of the Hubble deep field.

Since all the particles in a given region are tagged with the epoch
of formation, the mean - mass weighted -
age of any region is computable at any time.
The two youngest sets of galaxies with formation epochs
at $z<1.0$ (dotted and short-dashed curves)
are nearly unbiased at large smoothing
lengths, e.g., $r_{TH}\sim 15h^{-1}$Mpc
(above which the limited box size causes
the estimate to be inaccurate)
and have a bias
of $\sim 2.0$ at  $r_{TH}\sim 1h^{-1}$Mpc, below which
our simulation fails.
The next set of galaxies that formed at $z=1.0-1.3$ (long-dashed curve),
is more strongly biased than the two youngest sets of galaxies.
The oldest galaxies (dot-dashed curve)
are most strongly biased
with a bias $\sim 5.5$ at  $r_{TH}\sim 1h^{-1}$Mpc
and are still significantly biased
at large scales, e.g. $\sim 3.0$ at  $r_{TH}\sim 10h^{-1}$Mpc.
If we think of ellipticals as being the
stellar systems within which most star-formation was completed
earliest (\cf Searle \& Sargent 1972),
they should represent, in our terms,
the oldest systems.
We know these are very overabundant in the rare rich
clusters, i.e.,
in regions of very high 
overdensity,
and so are strongly biased even on large scales (Kaiser 1984).
We note from panel (a) that at the fiducial smoothing
scale of $8h^{-1}$Mpc the bias for all galaxies
is $1.35$, give $\sigma_{8,gal}=1.08$ (for $\sigma_{8,mass}=0.80$),
consistent with the observed value of 
$\sigma_8=1.20\pm 0.18$ (Loveday \etal 1996)
(Note that the observed value of $\sigma_8$ is for bright galaxies
in the APM survey; this is more appropriate for our comparison
because the computed galaxy fluctuation is mass-weighted, thus
heavily weighted by the most massive, bright galaxies).

Panel (b) shows directly the auto-correlation functions
for the stellar mass density of 
all galaxies, mass and the stellar mass densities
of galaxy subsets ordered by formation redshift;
the results are consistent with Panel (a) of Figure 2.
To summarize, 
all types of galaxies as well as all galaxies
are biased over matter on all scales with two separate trends:
1) at any epoch
older objects are more strongly clustered than are younger ones,
and 2) bias decreases with increasing scale.
Both trends are consistent with our earlier work (Cen \& Ostriker 1992).
A question has arisen as to whether antibias occurs in 
these simulations as it does in some quasi-analytic modeling.
The answer is ``yes" but for different reasons.
In dark matter simulations 
it is found (Kravtsov \& Klypin 1998)
that in dense regions
halos will merge, with the expectation that the imbedded
galaxies will merge as well,
reducing the galaxy number density of galaxies
(but not the light in galaxies).
This effect occurs as well in our simulations
but, as is evident from Figure 1, merger of halos 
does not necessarily lead to mergers of the dissipational galactic
components.
We find antibias for young systems in dense regions (Blanton \etal 1999b)
for the same reason that star-forming regions are rare
in real rich and dense clusters of galaxies:
the ambient very hot ($T\sim 10^8$Kelvin) gas
can not cool and thus can not accrete onto dense lumps of collisionless
stellar or dark matter.
This effect sets in as soon as the strong caustics form
in our simulations,
and considerably before the final accumulation of virialized clusters.

We also sorted our computed sample into four sets 
by the metallicity of their first generation stars
$(Z/Z_\odot=[<-1.69], [-1.69 \rightarrow -1.57], [-1.57 \rightarrow -1.48], [>-1.48])$.
The trends are comparable to those seen in Panels (a,b) of Figure 2
in the sense that
galactic regions with initial high metallicity are more strongly biased
than are those with initial low metallicity and
any set of objects has an increasing bias with decreasing scale.
This presumably reflects the physical situation that 
high density regions generally have higher metallicity.
However, the trend is substantially weaker
than that seen in subsets ordered
by their formation time.

In Figure 3 we compare the simulation results with observations.
In Panel (a) we select the second oldest quartile of objects
and designate it as ``elliptical galaxies",
whose correlation function
is shown along with the correlation
functions of observed elliptical galaxies
from two independent redshift surveys.
Panel (b) shows the correlation function
for the second youngest quartile of objects,
which we call ``spiral galaxies";
also shown are the correlation functions of 
observed spiral galaxies from redshift surveys.
Finally, 
in Panel (c) we show the correlation function
for the density in all objects and total mass in comparison with
observed correlations of all galaxies from 
three redshift surveys.
While the separate subsets (panels a,b)
produce results in good agreement with their 
observational
counterparts in the range of scales where our simulation results are 
most accurate ($1-10h^{-1}$Mpc),
the agreement between observations and simulations 
for all galaxies are not satisfactory.
The likely cause is that, while we are computing the correlation functions
of galaxy {\it mass} density, the observed correlation functions are
galaxy number weighted
and in high density regions the number density of objects
will decrease due to mergers while
leaving the mass density in galaxies invariant.
More robust is the trend
found in our computations that 
older galaxies tend to cluster more strongly 
than the average in a fashion which
is fully consistent with observations.

Figure 4 shows the void probability function
as a function of top-hat sphere radius.
The void probability function is defined
as the probability of a (top-hat) sphere
of the given radius having a density of one-fifth of
its corresponding global mean (Weinberg \& Cole 1992).
We see that the probability of having a void as large
as $10h^{-1}$Mpc in the mass distribution (dot-dashed curve)
is 0.005 (the limited box size
of the simulation does not allow a probability $<0.005$ to be
estimated).
This has been highlighted by Peebles (1993)
as a potential problem for CDM theories.
But the void probabilities
for various subsets of galaxies and all galaxies
are much larger, at $\sim 0.2-0.6$, at $r_{TH}=10h^{-1}$Mpc.
Here we see that the simulations quite naturally produce
a universe which over half of the volume is in voids
with radii $>5h^{-1}$Mpc,
while only about $10\%$ of the dark matter universe
is in similar voids.
Also shown as symbols are observations from Vogeley \etal (1994)
of the CfA survey for volume-limited samples with absolute
magnitude less than -19.5 (solid squares) and -20.0 (open squares),
respectively (fainter samples are not shown because our simulations
are likely to have underproduced a significant number of fainter galaxies).
Comparing with observations of Vogeley \etal (1994) we see 
that the real galaxies are indeed quite ``voidy", consistent
with the distribution of our simulated galaxies,
but inconsistent with 
the distribution of mass, implying the necessity of biased
galaxy formation.
Also consistent with observations are the common trend that
older galaxies are more voidy than younger ones,
consistent
with the fact that they tend to reside in high density regions
like rich clusters of galaxies.
In the underdense regions
examined in the analyses of voids,
our spatial resolution is adequate to identify
luminous galaxies.

Figure 5 shows peculiar velocity distribution of galaxy mass
at $z=0$ averaged over a top-hat smoothing scale of $1h^{-1}$Mpc,
a typical group scale.
Two things come to notice.
First, groups of younger galaxies tend to move more slowly than
clusters of older galaxies with the difference 
of the median values between the youngest quartile
and the oldest quartile being approximately 100km/s.
Second, averaging over all objects, stellar material
tend to move slightly
more slowly than does dark matter. 
The latter is equivalent to a small anti-velocity bias,
reflecting the gasdynamic effect,
while the former clearly indicates that
older galaxies tend to reside in higher density, deeper potential 
regions than do younger ones.
The much larger effect 
- that the small sacle velocity dispersion of spirals
is far less than the small scale velocity dispersion of ellipticals -
cannot be addressed in this work.

\subsection{Redshift Evolution of Bias of Galaxies}

Let us now study bias at high redshift.
Panel (c) of Figure 2 shows the bias 
as a function of radius
at four different redshifts.
It is seen that 
{\it bias is a monotonically increasing function of redshift}.
Second, 
{\it bias is a monotonically decreasing function of scale
at any redshift with the rate of change a function of smoothing
scale ($d b/d r_{cm}$) being nearly independent of redshift}.
We think that the increase of bias with redshift is due 
primarily to
the Kaiser (1984) effect.
It has also been found in the simulations of Katz \etal (1998).
Rarer and rarer events are required at higher and higher
redshifts to produce a galaxy of given mass.
Finally, we predict that galaxies should be
biased over mass by a factor $3-4$ on $\sim 10h^{-1}$Mpc scale at $z\sim 3$,
in excellent agreement with recent observations of
high redshift galaxies at $z\sim 3$
(Steidel \etal 1998), which show strong
concentrations of galaxies in narrow redshift bins
and imply a strong bias of approximately the same magnitude.
Note that our resolution at redshift
$z=3$ 
is $\sim 50h^{-1}$kpc and 
and $\sim 25h^{-1}$kpc in the two simulations

In addition we have plotted 
the auto-correlation functions
for all galaxy mass density and total mass density 
at various redshifts,
as a function of {\it comoving} separation.
We have found that, while
the correlation of mass grows with time as predicted
by linear theory,
the correlation function of galaxy density
as a function of {\it comoving} separation
evolves very weakly (due to the countervailing evolution of bias) 
in the pair separation range $r\ge 2.5h^{-1}$Mpc.
This was also found by Katz \etal (1998)
in a higher resolution simulation of a much smaller
box ($11.1h^{-1}$Mpc) which uses
the same galaxy identification scheme as we do.
Where our simulations overlap at redshift $z=3$
and separation $r=2h^{-1}$Mpc,
they obtain $\xi=0.3$ and we find $\xi=15.0$.
The very significant difference is probably due to the differences
in simulation box sizes.

In Figure 6 we address evolution in another way by examining 
the correlation length of the smoothed (mass-weighted) galaxy density
as a function of redshift.
Shown for comparison are the dark matter particle-particle (short-dashed
curve) and halo-halo (long-dashed curve) correlation lengths.
Also shown as symbols and 
a shaded region are available observations from various sources.
When needed, observations are converted
to be consistent with the adopted cosmological model.
We see that, as expected, the correlation function of mass
(dark matter particles) 
is a monotonically deceasing function of redshift:
gravitational instability (Peebles 1980) increases 
correlation with time.
On the other hand, the correlation of halos, selected by
identifying all regions (cells) at or above the virial density
at each redshift,
shows a stronger strength at all redshifts 
than that of galaxy density.
Furthermore, the halo correlation shows an interesting ``dip"
near $z\sim 0.5$, in contrast to the galaxy density correlation which
shows a weak peak at $z\sim 1.0$.
The differences could be explained in part by numerical resolution effect
and in part by real physical effects.
Relatively larger dark matter particle mass compared to
the baryonic mass resolution element in the simulation
preferentially biases against identification of less massive halos,
thus overestimates the correlation of halos at all redshifts.
The other effect is physical.
At lower redshift ($z\le 1.0$)
the galaxy formation preferentially avoids rich clusters
of galaxies, simply because these sites
are (as noted earlier)
too hot to permit proper cooling and gravitational
instabilities to occur effectively.
This explains the opposite trends in the halo and galaxy correlations
at $z<0.5-1.0$.

Kauffman \etal (1998) have also recently addressed
this question and find a drop in the (comoving) correlation length
at $z\sim 0.5$, very similar to what 
we found for dark matter halos,
but contrary to what we found
for our smoothed galaxy distribution.
Hence their result may be due to the method utilized,
which is based on
identifying N-body halos within which galaxies are expected to reside
rather than attempting a full physical simulation.
Thus, our result on the correlation evolution
for halos (long-dashed curve in Figure 6)
is consistent with both that of Colin \etal (1998) 
from pure N-body simulations and that of Kauffmann \etal (1994) of
gas-added halos in the general
trend that the halo correlation decreases from redshift $z\ge 3$ to
$z\sim 1$ and then rises towards zero redshift;
there is a minimum at $z\sim 1$.
We caution that the correlation
at high redshift may have been overestimated
in our simulation thanks to the limited resolution, which
preferentially selects out the most
massive halos that are most strongly clustered.
This indeed may explain the somewhat stronger correlation
found in our simulations than in Colin \etal at $z=3$.
At lower redshift this effect is progressively less significant.

Overall, the fit to observation is reasonably good, with
both roughly consistent with being flat in the range $z=0-3$.
However, one needs to keep in mind that our computed
correlation functions are 
weighted by the mass in galaxies not the number. 
Because of this, one expects to have better agreements between our results
and observations which look
at all galaxies and do not exclude high density regions,
if the cosmological model is right and the physical modeling
is appropriate.
Indeed, the top open square in Figure 6  
of Shepherd \etal (1997) data points,
which does not exclude clusters in the sample,
agrees much better with our results than the bottom open
square with an ``x" in the middle which does exclude clusters.
The same can be said about data point of Giavalisco \etal (1998; open circle
with an ``x" in the middle),
which has a fainter sample than that of Adelberger \etal (1998; open triangle).

\section{Conclusions}
We analyze a new physically based large ($100h^{-1}$Mpc)
numerical simulation of a plausible $\Lambda$CDM cosmological
model to quantify the relation expected between mass density and 
smoothed galaxy density.
The results, while consistent, as far as we know,
with existent observations 
are quite inconsistent with the simplest model,
where the ratio of galaxy fluctuations to mass fluctuations is a number, $b$,
called bias.
This ratio, in principle, could depend on the physical
scale being considered,
the age or metallicity of the galaxies being studied or the epoch
at which the analysis is made.
We find in our simulation that bias increases with decreasing
scale, with increasing galactic age or metallicity and
with increasing redshift.
Looking at the average galaxy mass density at redshift zero our
result, $b=1.35$
is consistent with the observed value $1.50\pm 0.19$ (Loveday \etal 1996;
the quoted observational value is for bright galaxies in the APM survey,
which is appropriate for comparison with our model because the computed
galaxy fluctuation is mass-weighted, not number-weighted, thus
heavily weighted by the most massive, bright galaxies).
All of the proposed dependencies can be tested
by comparing to observations and if confirmed,
used to interpret large-scale galactic surveys now underway.
The surprising emptiness of voids also follows naturally from this physical
treatment.
We also predict that, due to an apparent coincidence between opposing
trends,
the strength of the spatial correlation between pairs of galaxies (as 
a function of {\it comoving} separation)
should be essentially independent of redshift from $z=0$ to $z=3$.
Here we differ from the Kauffmann \etal (1998) result,
thus providing an opportunity
to distinguish between the accuracy of the two methods
currently being utilized for computing large-scale
structure properties from {\it ab initio} methods.
Lastly we find that the perturbations on the Hubble flow for young
(e.g., late type spiral or irregular)
galaxies should be less than for early type systems;
this is a testable proposition.

Much of the complexity and stochasticity of
the reported results is due, we believe,
to the fact that the conventional treatments (including this one)
are not examining all of the right variables.
Blanton \etal (1999a,b) show that galaxy density is
a function not just of mass density, but also of potential (or 
velocity dispersion or temperature) in the region under
examination.
We will return to this in future work with an analysis of the
galaxy data  using more appropriate variables.

\acknowledgments
Discussions with Michael Blanton,
Guinevere Kauffmann, Jim Peebles, Michael Strauss  and
Michael Vogeley are gratefully acknowledged as
are comments by a referee which enabled us to substantially
clarify our presentation.
We thank Michael Vogeley for providing the
observations concerning void probability functions and correlation functions
from the CfA Survey,
Douglas Tucker for providing the
correlation function data from Las Campanas Redshift Survey 
and 
Jon Loveday for providing the
correlation function data from APM Survey.
The work is supported in part
by grants NAG5-2759 and AST93-18185, ASC97-40300.

\appendix
\section{Effect of Finite Numerical Resolution}

It is important to understand how the finite numerical
resolution might have affected our computed results.
We show in Figure 7 results from both $L=100h^{-1}$Mpc and 
$L=50h^{-1}$Mpc boxes.
We smoothed the galaxy and dark matter density fields by 
a Gaussian window of the same radius $0.5h^{-1}$Mpc in both boxes.
Panel (a) of Figure 7 shows the bias (defined by equation 1)
from the two boxes and panel (b) shows
the square root of the ratio of galaxy correlation function to
total matter correlation function.
We see that two boxes
give comparable results with a maximum difference
less than $9\%$ (at $r_{TH}=1h^{-1}$Mpc, below which
no comparison is made; the oscillations at $r_{sp}> 5h^{-1}$Mpc
on panel (b) are just noise).
We therefore conclude that,
for the range of scale of interest $r\ge 1h^{-1}$Mpc,
our finite numerical resolution does not significantly
affect the results, at least for the quantities examined
in this paper, namely bias 
of the smoothed galaxy density
and correlation function of this same
function.

\newpage
\figcaption[FLENAME]{
The top (large) panel shows the dark matter density 
for a random slice of size $100\times 100h^{-2}$Mpc$^2$
and thickness of three cells ($586h^{-1}$kpc).
Each small tickmark of the top panel has a size of $6.25h^{-1}$Mpc.
The three bottom pairs of panels show three galaxy groups in this slice
in an enlarged display; two separate panels are shown for each selected
galaxies with the top panel showing the galaxy density contour
and the bottom panel showing the dark matter density contour.
Each tickmark for the bottom panels 
has a size of $391h^{-1}$kpc, corresponding to two cells,
which is approximately the resolution of the code
(Cen \& Ostriker 1999).
\label{fig1}}

\figcaption[FLENAME]{
Panel (a) shows the bias of galaxies
for four sets ($4\times 25\%$ in mass) 
of galaxies ordered by
formation time [z=(0.0-0.8), (0.8-1.0), (1.0-1.3), (1.3-10.0)],
as well as all galaxies at $z=0$.
Panel (b) shows the auto-correlation functions
for all galaxies, mass, galaxy subsets sorted by formation epoch at zero 
redshift.
Panel (c) shows the bias as a function of radius
at four different redshifts.
Panel (d) shows the auto-correlation functions at various epochs
for all galaxies and mass.
\label{fig2}}

\figcaption[FLENAME]{
Panel (a) shows the auto-correlation function for the
early type galaxies from the simulation (long-dashed curve with 
$1\sigma$ errorbars) at zero redshift;
the second oldest quartile of galaxies is used here.
Also shown in Panel (a) are observations
for ellipticals 
from an analysis of galaxies in the Perseus-Pisces region 
(solid dots; Guzzo \etal 1997)
and from the APM survey (solid triangles with $1 \sigma$ errorbars;
Loveday \etal 1995).
Panel (b) shows the auto-correlation function
for the late type galaxies from the simulation (dashed curve with 
$1\sigma$ errorbars) at zero redshift;
the second youngest quartile of galaxies is shown here.
The symbols in Panel (b) are observed correlations 
of spirals (open circles; Guzzo \etal 1997) in the Perseus-Pisces region
and from the APM survey (open triangles with $1 \sigma$ errorbars;
Loveday \etal 1995).
Panel (c) shows the auto-correlation function
for all galaxies (solid curve) and total mass (dot-dashed curve)
at zero redshift.
The symbols in Panel (c) are observed correlation functions
of galaxies from various surveys:
open squares from the Las Companas Redshift survey 
(Tucker \etal 1997),
filled squares from the APM survey (Loveday \etal 1995)
and solid dots from the CfA survey (Vogeley \etal 1993).
\label{fig2}}

\figcaption[FLENAME]{
Figure 4 shows the void probability function
as a function of top-hat sphere radius
for all mass, all stellar mass and three subsets of stellar mass
($25\%$, $50\%$ and $75\%$ oldest stellar mass).
The void probability function is defined
the probability of a sphere
of the given radius having a density of one-fifth of
its corresponding global mean (Weinberg \& Cole 1992).
Also shown as symbols are observations from Vogeley \etal (1994)
of the CfA survey for volume-limited samples with absolute
magnitude less than -19.5 (solid squares) and -20.0 (open squares),
respectively.
\label{fig4}}

\figcaption[FLENAME]{
Figure 5 shows peculiar velocity distributions of various sets
of galaxies and dark matter
at $z=0$ averaged over a top-hat smoothing scale of $1h^{-1}$Mpc. 
The vertical lines indicate the respective
medians of the various sets of objects.
\label{fig5}}

\figcaption[FLENAME]{
Figure 6 shows 
the correlation length of all galaxies (mass-weighted)
as a function of redshift.
Also shown are available observations from various sources:
the solid dot is from APM observations (Loveday \etal 1995)
and Las Companas Redshift survey (Lin 1995) at redshift zero;
the solid square is the Autofib survey by Cole \etal (1994) at $\bar z=0.16$;
the open squares are form the Canadian Network for Observational
Cosmology cluster survey by Shepherd \etal (1997)
where the lower square with an ``x" in the middle
is from extended field sample
and the upper square is when the redshift subsample is considered,
at $\bar z=0.37$;
the cross-shaded region indicates
the result from CNOC2 by Carlberg \etal (1998)
at $\bar z=0.0-0.7$;
the open circle with an ``x" in the middle
is from Giavalisco \etal (1998) from analysis of Lyman Break
Galaxies at $\bar z=3.07$;
the open triangle is from Adelberger \etal (1998) from 
counts-in-cells analysis of Lyman Break Galaxies at $\bar z=3.07$;
the filled triangle is from Postman \etal (1998) from 
an angular correlation analysis based on 710,000 galaxies with $I_{AB}<24$
from a survey of $4\times 4$ square degree, contiguous sky.
Also shown for comparison are the dark matter particle-particle (short-dashed
curve) and halo-halo (long-dashed curve) correlation lengths.
\label{fig6}}

\figcaption[FLENAME]{
Panel (a) shows the bias (defined by equation 1)
from the two boxes with size $L=(100,50)h^{-1}$Mpc
and panel (b) shows
the square root of the ratio of galaxy correlation function to
total matter correlation function.
The galaxy and dark matter density fields in both boxes
are smoothed by a Gaussian window of radius $0.5h^{-1}$Mpc.
The errorbars shown are $1 \sigma$.
\label{fig7}}


\clearpage

\begin{thebibliography}{DUM}
\bibitem[Abel \etal 1998]{a98} Abel, T., Anninos, P., Norman, M.L., \& Zhang, Y. 1998, ApJ, 508, 518
\bibitem[Adelberger \etal 1998]{g98} Adelberger, K., Steidel, C.C., Giavalisco, M., Dickinson, M.E., Pettini, M., \& Kellogg, M. 1998,  preprint, astro-ph/9804236
\bibitem[Bahcall 1988]{b88} Bahcall, N.A. 1988, ARAA, 26, 631
\bibitem[Blanton \etal 1999a]{bcos99} Blanton, M., Cen, R., Ostriker, J.P., \& Strauss, M.A. 1999a, ApJ, 522, 590
\bibitem[Blanton \etal 1999b]{bcos99} Blanton, M., Cen, R., Ostriker, J.P., Strauss, M.A., \& Tegmark, M. 1999b, preprint, astro-ph/9903165
\bibitem[Carlberg \etal 1998]{carl98} Carlberg, R.G., \etal, preprint, astro-ph/9805131
\bibitem[Cen 1992]{c92} Cen, R. 1992, \apjs, 78, 341
\bibitem[Cen \& Ostriker 1992]{co92} Cen, R., \& Ostriker, J.P. 1992, \apj, 399, L113
\bibitem[Cen \& Ostriker 1993a]{co93a} Cen, R., \& Ostriker, J.P. 1993a, \apj, 417, 404
\bibitem[Cen \& Ostriker 1993b]{co93b} Cen, R., \& Ostriker, J.P. 1993b, \apj, 417, 415
\bibitem[Cen \& Ostriker 1999]{co99} Cen, R., \& Ostriker, J.P. 1999, \apj, in press
\bibitem[Cen \etal 1995]{ckor95} Cen, R., Kang, H., Ostriker, J.P., \& Ryu, D. 1995, \apj, 451, 436
\bibitem[Cen \etal 1998]{cpmo98} Cen, R., Phelps, S., Miralda-Escud\'e, J., \& Ostriker, J.P. 1998, \apj, in press
\bibitem[Cole \etal 1994a]{cebc94} Cole, S., Ellis, R., Broadhurst, T., \& Colless, M. 1994, \mnras, 267, 541
\bibitem[Cole \etal 1994b]{cafnz94} Cole, S., Aragon-Salamanca, Frenk, C.S., Navarro, J., \& Zepf, S.E. 1994, \mnras, 271, 781
\bibitem[Colin \etal 1998]{c98} Colin, P., Klypin, A.A., Kravtsov, A.V., \& Khokhlov, A.M. 1998, preprint, astro-ph/9809202
\bibitem[Davis \& Peebles 1983]{dp83} Davis, M., \& Peebles, P.J.E. 1983, \apj, 267, 465
\bibitem[Davis \etal 1985]{defw85} Davis, M., Efstathiou, G., Frenk, C.S., \& White, S.D.M. 1985, \apj, 292, 371
\bibitem[Eggen, Lynden-Bell, \& Sandage 1962]{els62} Eggen, O.J., Lynden-Bell, D., \& Sandage, A.R. 1962, ApJ, 136, 748
\bibitem[Eke, Cole, \& Frenk 1996]{ecf96} Eke, V.R., Cole, S., \& Frenk, C.S. 1996, \mnras, 282, 263
\bibitem[Evrard \etal 1994]{esd94} Evrard, A.E., Summers, F.J., \& Davis, M. 1994, \apj 422, 11
\bibitem[Ferland 1994]{f94} Ferland, G.J. 1994, preprint
\bibitem[Frenk \etal 1985]{fwed85} Frenk, C.S., White, S.D.M., Efstathiou, G., \& Davis, M. 1985, \nat, 371, 595
\bibitem[Giavalisco \etal 1998]{gi98} Giavalisco, M., Steidel, C.C., Adelberger, K.L., Dickinson, M.E., Pettini, M., \& Kellogg, M. 1998,  preprint, astro-ph/9802318
\bibitem[Gnedin \& Ostriker 1997]{go97} Gnedin, N.Y., \& Ostriker, J.P. 1997, 486, 581
\bibitem[Guzzo \etal 1997]{g97} Guzzo, L., Strauss, M.A., Fisher, K.B., Giovanelli, R., \& Haynes, M.P. 1997, \apj, 489, 37
\bibitem[Harten 1983]{h83} Harten, A. 1983, J. Comp. Phys., 49, 357
\bibitem[Kaiser 1984]{k84} Kaiser, N. 1984, \apj, 284, L9
\bibitem[Katz, Hernquist, \& Weinberg 1992]{khw92} Katz, N., Hernquist, L., \& Weinberg, D.H 1992, \apj, 399, L109
\bibitem[Katz, Hernquist, \& Weinberg 1998]{khw98} Katz, N., Hernquist, L., \& Weinberg, D.H 1998, preprint, astro-ph/9806257
\bibitem[Katz, Weinberg, \& Hernquist 1996]{kwh96} Katz, N., Weinberg, D.H., \& Hernquist, L. 1996, \apjs, 105, 19
\bibitem[Kauffmann, Nusser, \& Steinmetz 1997]{kns97} Kauffmann, G., Nusser, A., \& Steinmetz, M. 1997, \mnras, 286, 795
\bibitem[Kauffmann \etal 1998]{k98} Kauffmann, G., Colberg, J.M., Diaferio, A., \& White, S.D.M. 1998, preprint, astro-ph/9809168
\bibitem[Kennicutt 1989]{k89} Kennicutt, R.C. 1989, ApJ, 344, 685
\bibitem[Krauss \& Turner 1995]{kt95} Krauss, L., \& Turner, M.S. 1995, Gen. Rel. Grav., 27, 1137
\bibitem[Kravtsov \& Klypin 1998]{kk95} Kravtsov, A., \& Klypin, A. 1998, preprint, astro-ph/9812311
\bibitem[Lin 1995]{lin95} Lin, H. 1995, Ph.D. thesis, Harvard University
\bibitem[Loveday \etal 1995]{l95} Loveday, J., Maddox, S.J., Efstathiou, G., \& Peterson, B.A. 1995, \apj, 442, 457
\bibitem[Loveday \etal 1996]{l96} Loveday, J., Efstathiou, G.,  Maddox, S.J., \& Peterson, B.A. 1996, \apj, 468, 1
\bibitem[Madau]{Madau97} Madau, P., 1997, in AIP Conf. Proc. 393, Star Formation Near and Far, ed. S.~S.Holt~\& G.~L.~Mundy (New York:AIP), 481
\bibitem[Ostriker \& Steinhardt 1995]{os95} Ostriker, J.P., \& Steinhardt, P. 1995, \nat, 377, 600
\bibitem[Peebles 1972]{p72} Peeles, P.J.E 1972, Physical Cosmology (Princeton: Princeton University Press)
\bibitem[Peebles 1980]{p80} Peeles, P.J.E 1980, The Large-Scale Structure of the Universe (Princeton: Princeton University Press)
\bibitem[Pen 1997]{p97} Pen, U.-L. 1997, preprint, astro-ph/9711180
\bibitem[Postman \etal 1998]{p98} Postman, M., Lauer, T.R., Szapudi, I., \&, Oegerle, W.e 1998, preprint, astro-ph/9804141
\bibitem[Raymond \& Smith 1977]{rs77} Raymond, J.C., \& Smith, B.W. 1977, ApJS, 35, 419 
\bibitem[Ryu \etal 1993]{rokc93} Ryu, D., Ostriker, J. P., Kang, H., \& Cen, R.  1993, \apj, 414, 1 
\bibitem[Salaris, Degl'Innocenti, \& Weiss 1997]{sdw97} Salaris, M., Degl'Innocenti, S., \& Weiss, A. 1997, \apj, 479, 665
\bibitem[Scalo 1986]{s86} Scalo, J.M. 1986, Fund. Cosmic Phys., 11, 1
\bibitem[Schmidt 1959]{s59} Schmidt, M. 1959, ApJ, 129, 243
\bibitem[Scherrer \& Weinberg 1997]{sw97} Scherrer, R.J., \& Weinberg, D.H. 1997, preprint, astro-ph
\bibitem[Serle \& Sargent 1972]{ss72} Serle, L., \& Sargent, W.L.W. 1972, \apj, 173, 25
\bibitem[Shepherd \etal 1997]{s97} Shepherd, C.W., Carlberg, R.G., Yee, H.K.C., \& Ellingson, E. 1997, \apj, 479, 82
\bibitem[Steidel \etal 1998]{s98} Steidel, C.C., Adelberger, K.L., Dickinson, M., Pettini, M., Giavalisco, M., \& Kellogg, M. 1998, \apj, 492, 428 
\bibitem[Steinmetz 1996]{s96} Steinmetz, M. 1996, \mnras, 278, 1005
\bibitem[Trimble 1997]{t97} Trimble, V. 1997, ``The Extragalactic Distance Scale" ed.  M. Livio, M. Donahue \& N. Panagia, p407
\bibitem[Tytler, Fan, \& Burles 1996]{tfb96} Tytler, D., Fan, X.-M., \& Burles, S. 1996, \nat 381, 207
\bibitem[Weinberg \& Cole 1992]{wc92} Weinberg, D.H., \& Cole, S. 1992, \mnras, 259, 652
\bibitem[Weinberg, Hernquist, \& Katz 1997]{whk97} Weinberg, D.H., Hernquist, L., \& Katz, N. 1997, \apj, 477, 8
\bibitem[White \etal 1987]{wdef93} White, S.D.M., Davis, M., Efstathiou, G., \& Frenk, C.S. 1987, \nat, 330, 451
\bibitem[White \etal 1993]{wef93} White, S.D.M., Efstathiou, G., \& Frenk, C.S. 1993, \mnras, 262, 1023
\bibitem[White \etal 1993]{wnef93} White, S.D.M, Navarro, J., Evrard, A.E., \& Frenk, C.S. 1993, \nat, 366, 429
\bibitem[Vogeley 1993]{v93} Vogeley, M.S. 1993, Ph.D Thesis, Harvard University
\bibitem[Vogeley \etal 1994]{v94} Vogeley, M.S., Geller, M.J., Park, C., \& Huchra, J.P. 1994, AJ, 108, 745
\end{thebibliography}
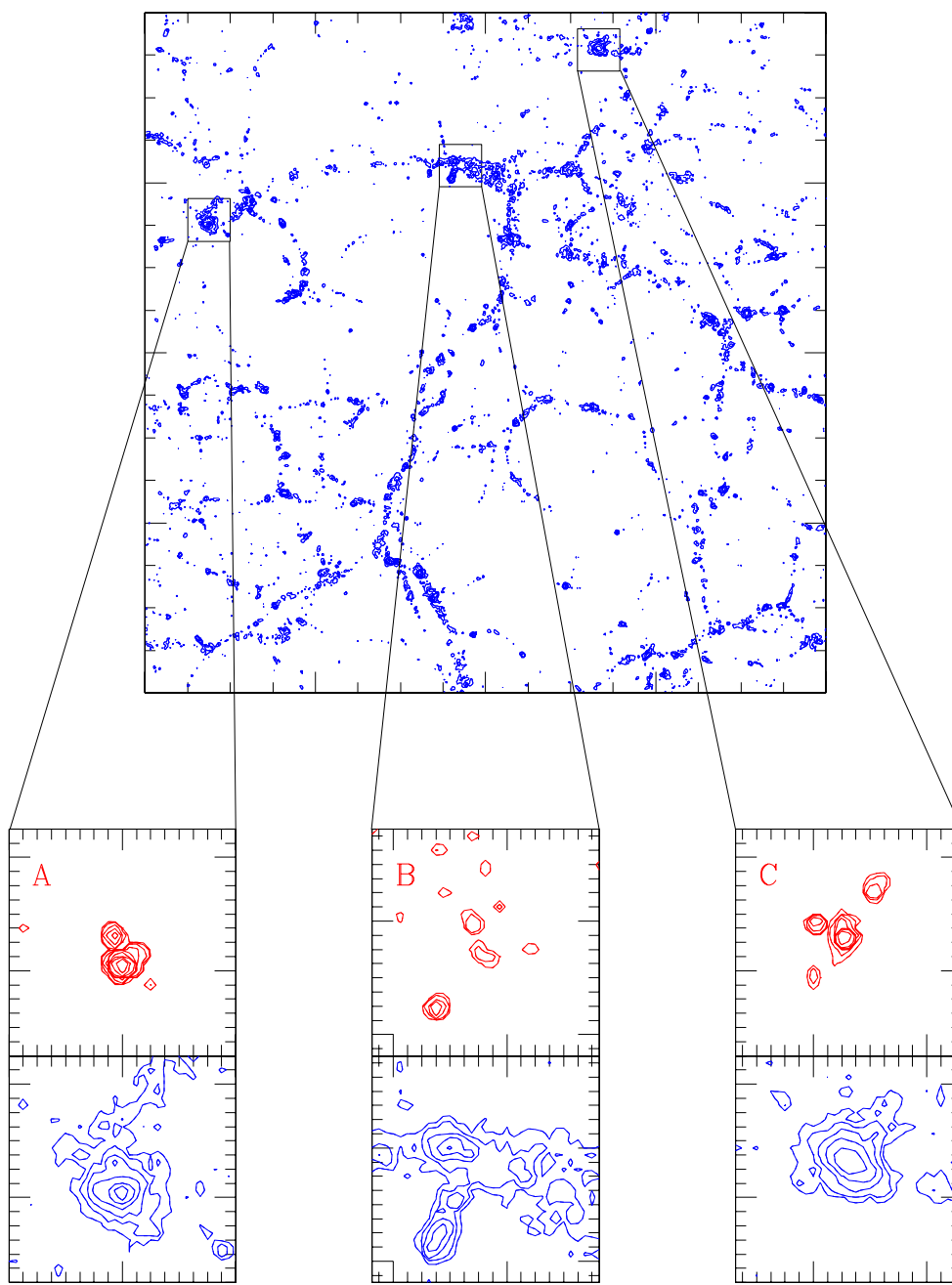
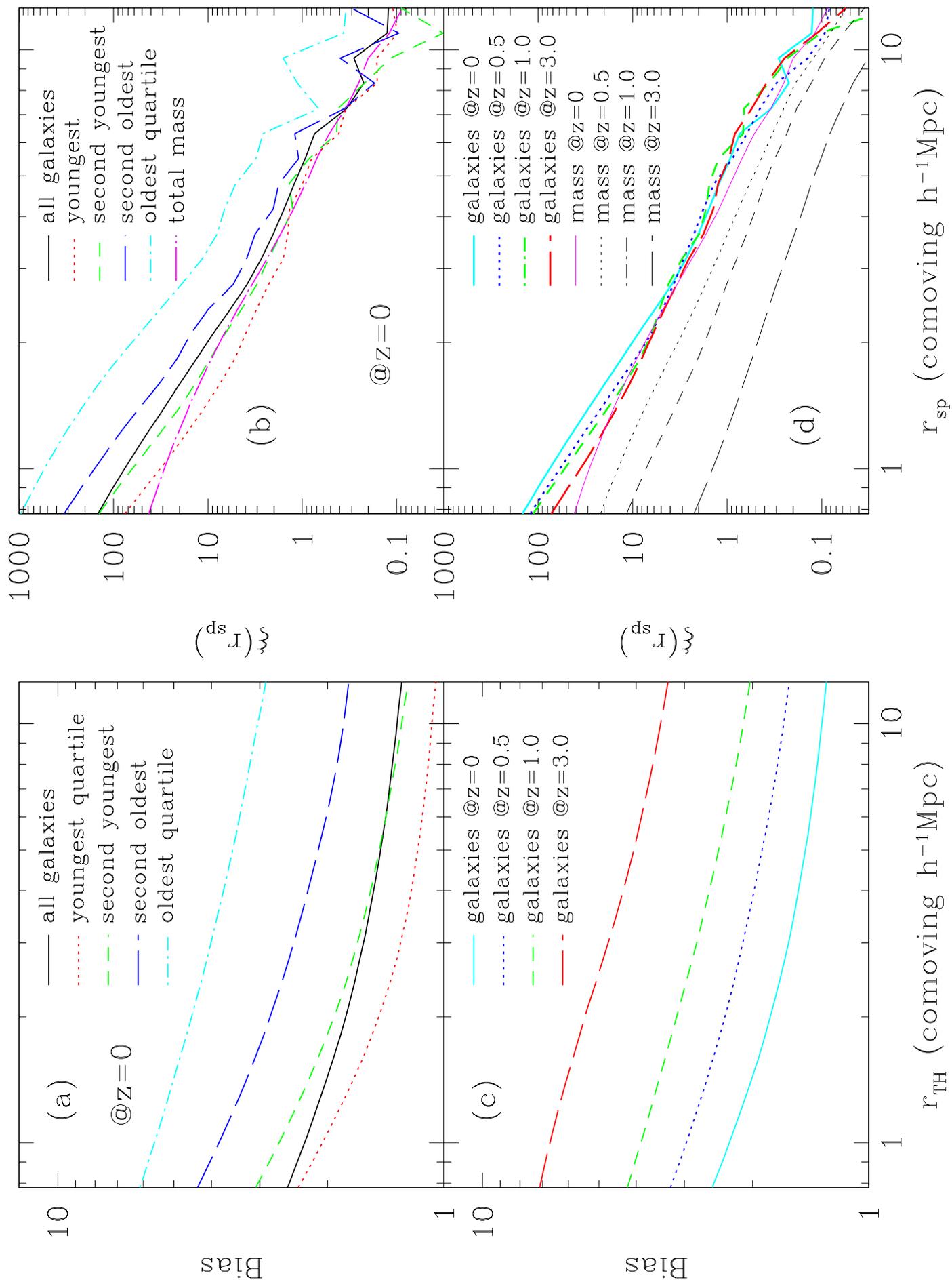
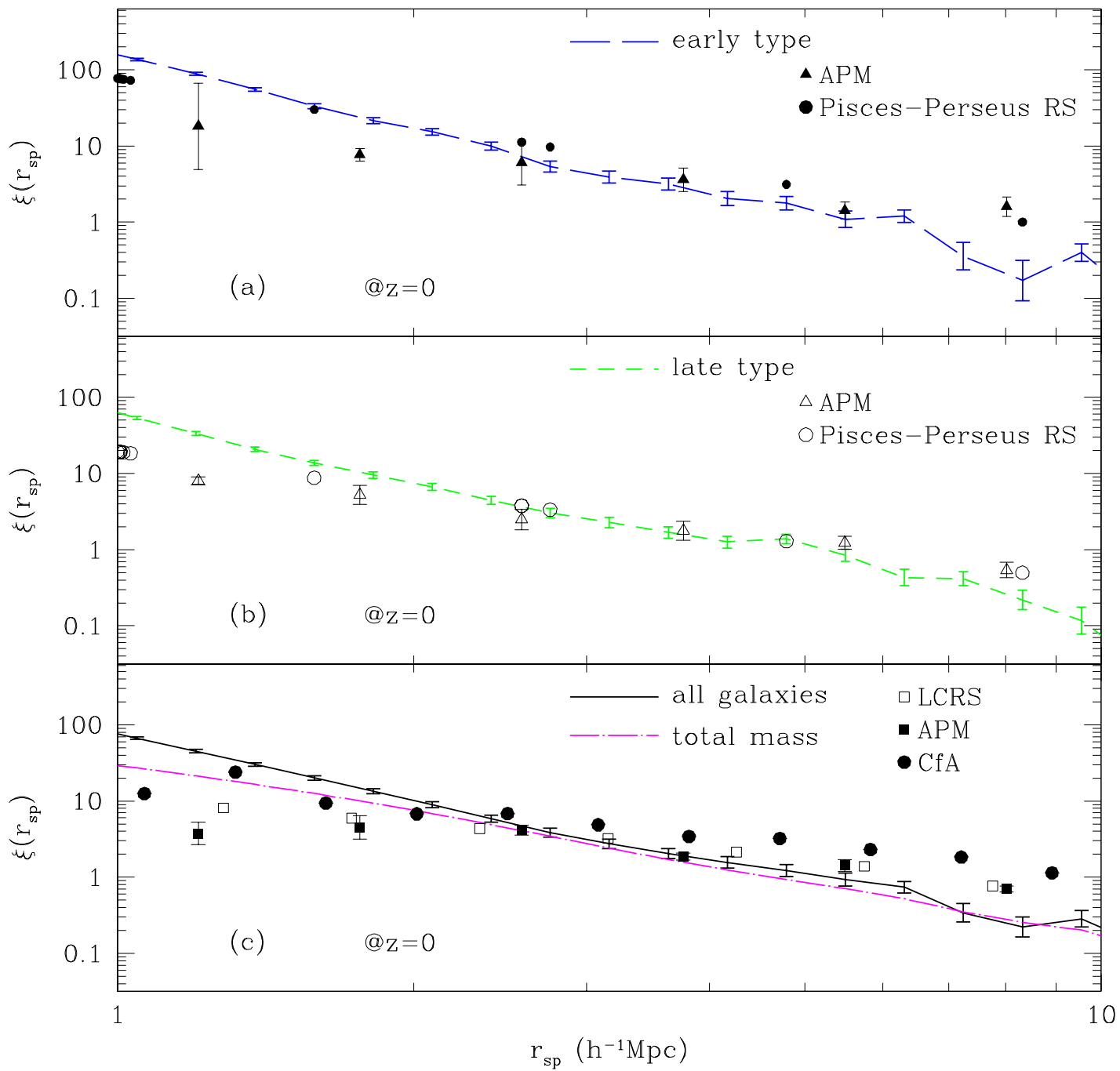
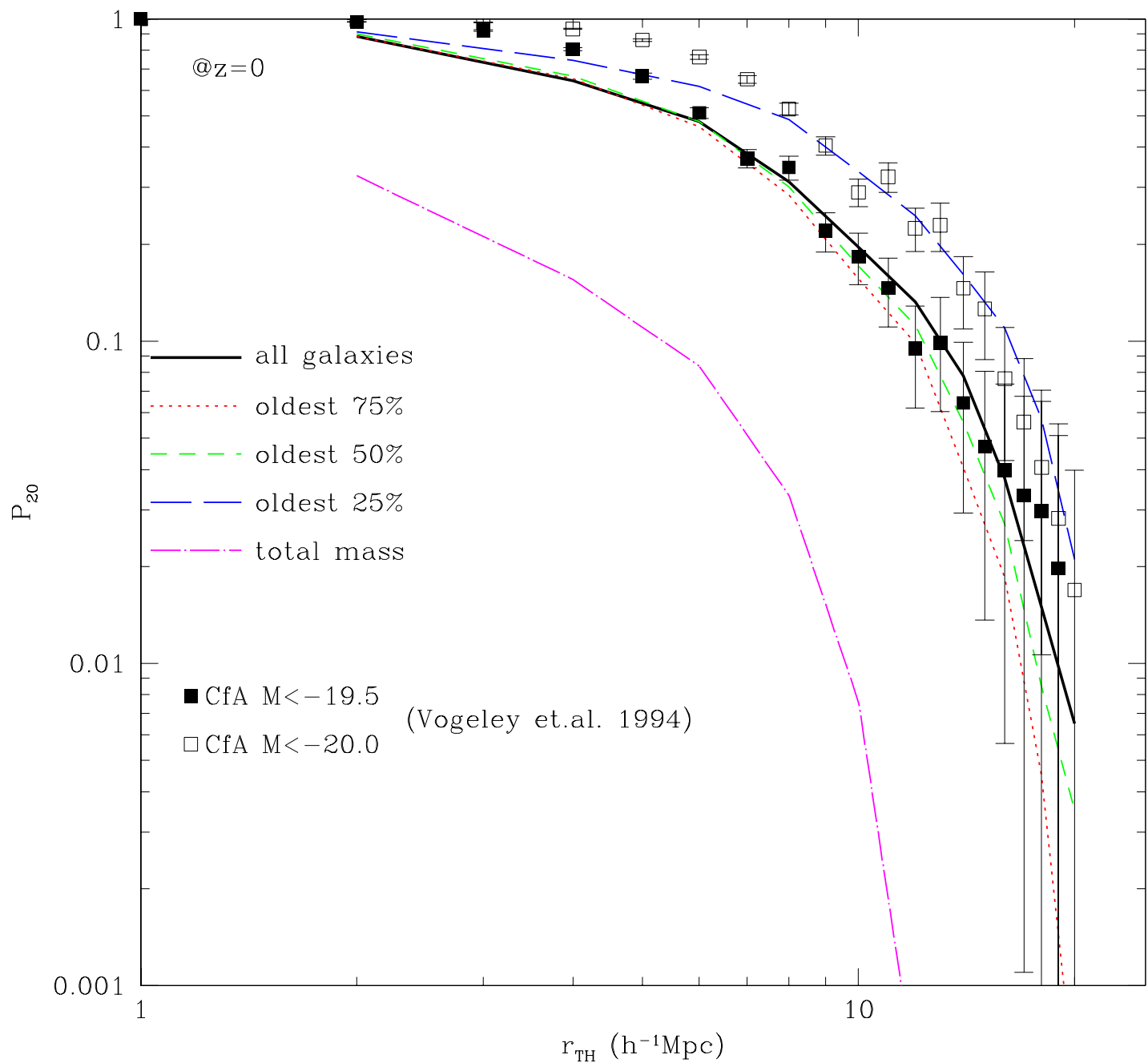
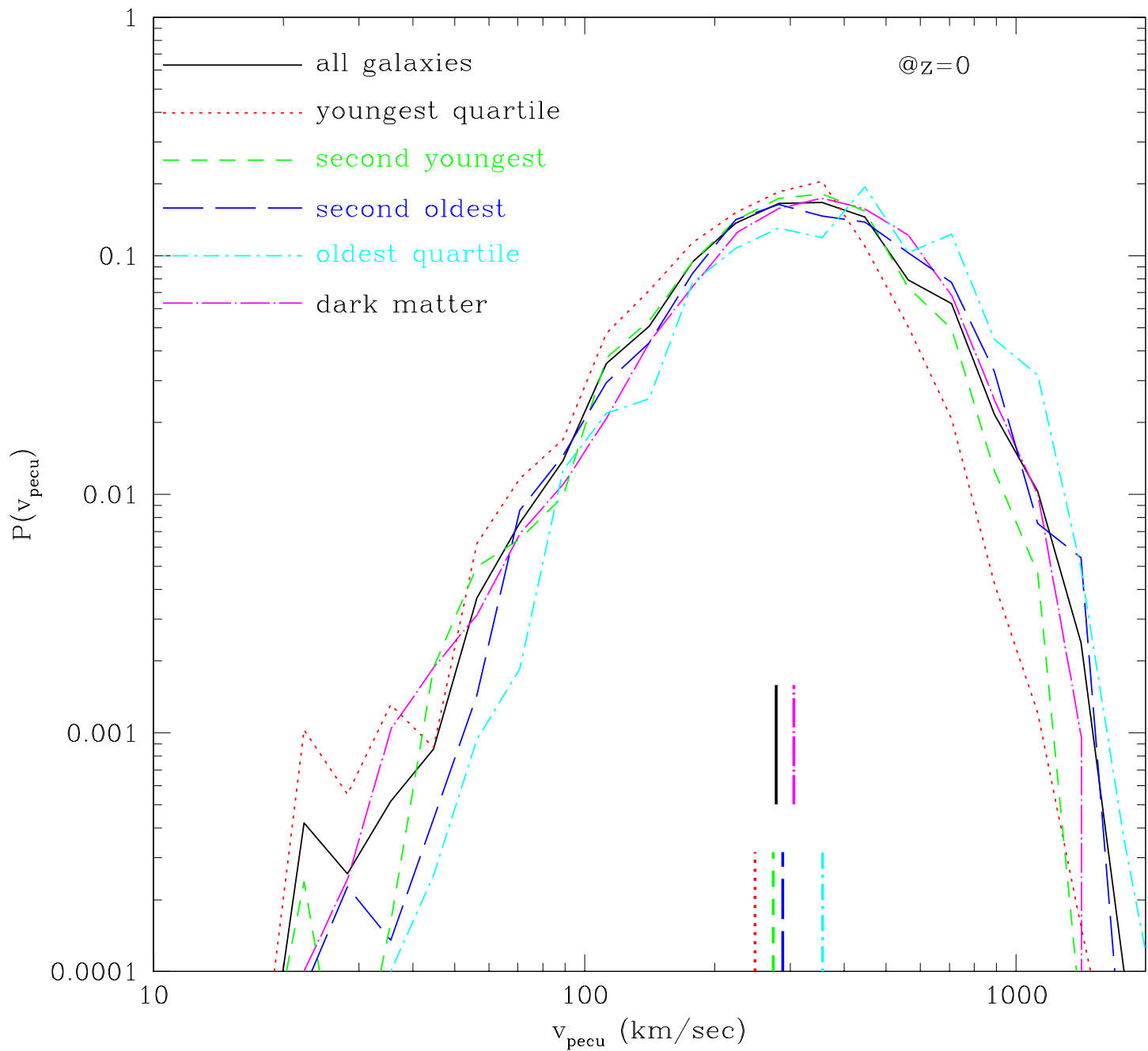
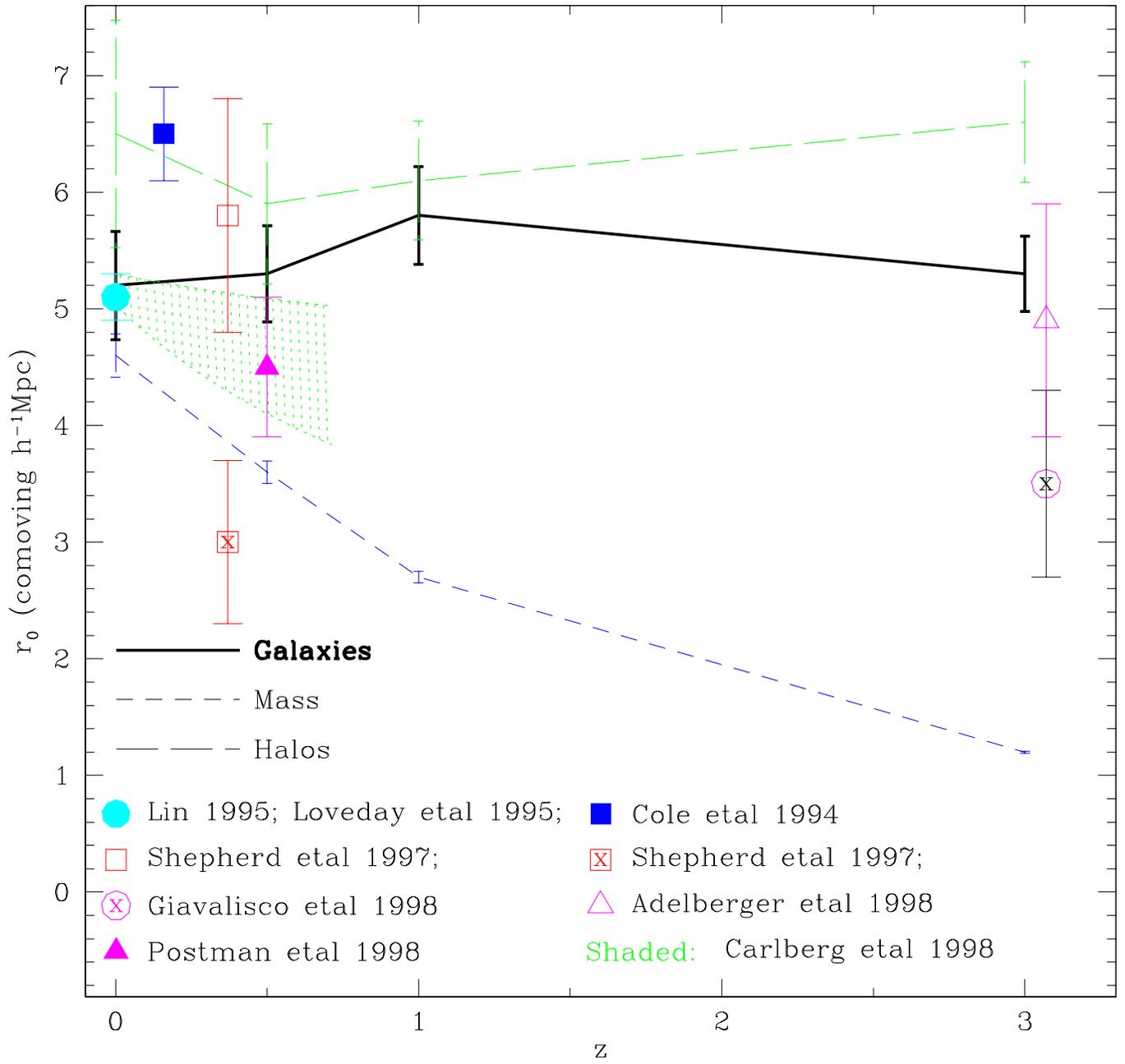
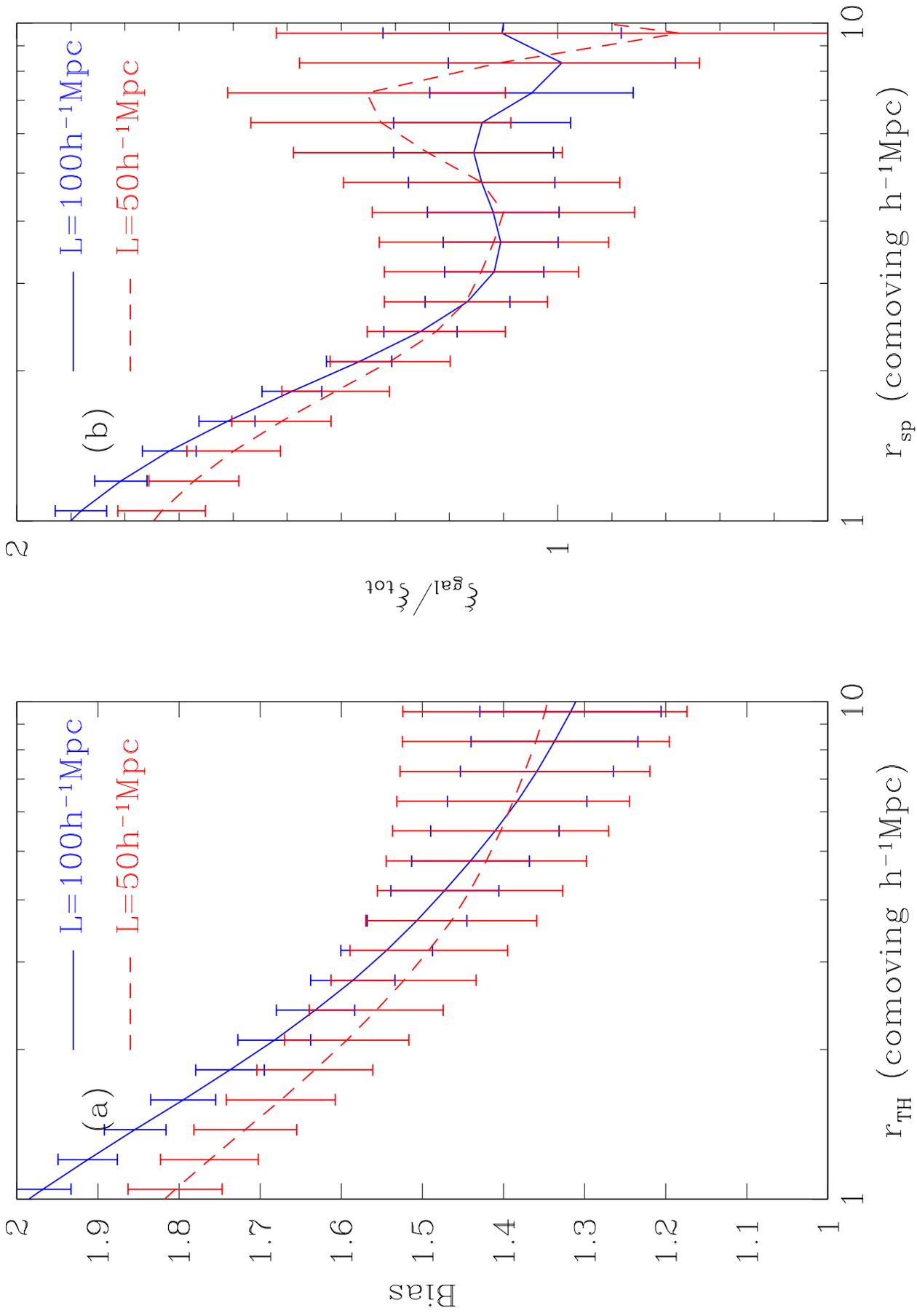
\end{document}